\documentclass[12pt,english]{article}
\usepackage[T1]{fontenc}
\usepackage[latin1]{inputenc}
\usepackage{amsmath}
\usepackage{amssymb}
\usepackage{mathrsfs}

\makeatletter

\catcode`@=12 \topmargin -0.5in \evensidemargin 0.0in
\usepackage{babel}
\makeatother
\newcommand {\be}{\begin{equation}}
\newcommand {\ee}{\end{equation}}
\newcommand {\ba}{\begin{eqnarray}}
\newcommand {\ea}{\end{eqnarray}}
\begin{document}
\thispagestyle{empty}
\begin{flushright}
IPM/P-2007/067\\
\today
\end{flushright}

\mbox{} \vspace{0.75in}

\begin{center}\textbf{\large A new class of invariants in the lepton sector}\\
 \vspace{0.5in} \textbf{$\textrm{Arman Esmaili}^{\dag\S}$\footnote{arman@mail.ipm.ir} and $\textrm{Yasaman Farzan}^{\S}$\footnote{yasaman@theory.ipm.ac.ir}}\\
 \vspace{0.2in} \textsl{${}^\dag$Department of Physics, Sharif University of Technology\\P.O.Box 11365-8639, Tehran, IRAN}\\
 \vspace{0.2in}\textsl{${}^\S$Institute for Studies in Theoretical Physics and Mathematics (IPM)\\P.O.Box 19395-5531, Tehran, IRAN}\\
 \vspace{.75in}\end{center}

\baselineskip 20pt
\begin{abstract}
We construct a new set of combinations from the mass matrices of
the charged leptons and neutrinos that are invariant under basis
transformation, hereafter {\it the} invariants. We use these
invariants to study various symmetries and neutrino mass textures
in a basis independent way. In particular, we show that by using
these invariants the ansatz such as $\mu-\tau$ exchange and
reflection symmetries, various texture zeros and flavor symmetries
can be expressed in a general basis.



\end{abstract}
\section{Introduction}
There is a consensus among  neutrino physicists that the recent
neutrino data from the solar and atmospheric neutrino observations
\cite{global}, the KamLAND reactor experiment \cite{kamland}  and
the long baseline experiments \cite{kek} can be explained only
through neutrino oscillation. As is well-known the oscillation
scenario is based on the fact that the neutrinos are massive and
do mix. The three by three mass matrix of the active neutrinos,
$m_\nu$, introduces new parameters to the standard model. Among
these parameters, six parameters in principle show up in the
neutrino oscillation probabilities: two mass-square splittings,
three mixing angles and one Dirac CP-violating phase. To this
list, one should add the mass scale of neutrinos ({\it i.e.,} the
mass of the lightest neutrino) which does not appear in the
oscillation probability and has to be derived in other types of
experiments such as the beta decay experiments or the cosmological
observations.

So far the nature of neutrinos (Majorana vs. Dirac) is not known.
However, the majority of neutrino mass models predict a Majorana
type neutrino mass matrix at low energies. Throughout this paper,
we shall assume that neutrinos are of Majorana type which implies
that their mass matrix is symmetric. If neutrinos are of Majorana
type, in addition to the above parameters the mass matrix will
contain two more physical degrees of freedom: two more
CP-violating phases which are called Majorana phases. These two
phases do not appear in the oscillation probabilities. Even if
neutrinos are proved to be of Majorana type, extracting the
Majorana phases is going to be quite challenging if possible at
all \cite{majorana-phases}. Moreover, only a combination of the
phases can be measured. That is, separately extracting each of the
Majorana phases will not be possible with the present methods.

Sources of CP-violation are associated with the phases of the
neutrino mass matrix; however, one has to be aware that by
rephasing the neutrino fields the phases of the elements of the
mass matrix also change. Since the seminal work by Jarlskog
\cite{Jarlskog}, defining invariants under field rephasing and
basis transformation has proved to be very useful for studying the
CP-violation in both quark and lepton sectors. An incomplete list
of the papers that have attempted to study the CP-violation in the
lepton sector by defining invariants is
\cite{kusenko,branco,us,sarkar,branco1}. As is well-known, because
of the presence of the extra CP-violating phases, the number of
the independent invariants is more than what we have in the quark
sector (or for the case of Dirac neutrinos). Recently, the
necessary and sufficient conditions for CP-violation has been
systematically formulated in terms of the rephasing invariants in
the mass basis of the charged leptons \cite{us,sarkar}. It is also
possible to study the CP-violation in terms of the combinations of
neutrino and charged lepton mass matrices that are invariant under
general basis transformation \cite{branco}. In this paper, we
introduce a new class of invariants under general basis
transformation. As we shall see, this new set of invariants is
very helpful for studying the symmetries of the neutrino mass
matrix.

Among the nine parameters of the neutrino mass matrix, the two
mass-square splittings and two of mixing angles are so far
measured. There are various running and planned experiments as
well as proposals to measure the remaining parameters. However, as
alluded to before, even in the most optimistic case, with the
present experiments and proposals, we will not be able to extract
all the neutrino parameters \cite{majorana-phases}. Motivated by
this fact various theoretical conjectures have been made to
reconstruct the neutrino mass matrix. Most of these conjectures
are based on symmetries that are apparent only in a particular
basis. Examples are texture zeros, $\mu-\tau$ exchange and
reflection symmetries \cite{us,Harrison} and various flavor
symmetries \cite{Binetruy}. All these symmetries are defined in
the mass basis of the charged leptons. Using the invariants
defined in the present paper, we can formulate these symmetries in
a basis-independent way.

This paper is organized as follows. In sect.~\ref{tarif}, we
introduce a new class of combinations that are invariant under
general basis transformation. In sect.~\ref{projections}, we use
the invariants for formulating the symmetries of the neutrino mass
matrix in a basis-independent way. We specially discuss the
$\mu-\tau$ exchange and reflection symmetries, flavor symmetry
conserving $L_\mu-L_\tau$ charge and texture zero ansatz. A
summary of results is given in sect. \ref{summary}.

\section{New class of invariants \label{tarif}}

Consider the following transformation on the charged leptons and
neutrinos:\ba \label{transformation1}
  \ell_{L\alpha} & \rightarrow& U_{\alpha\beta}\,\ell_{L\beta}\,,\nonumber \\
  \nu_{L\alpha} &\rightarrow& U_{\alpha\beta}\,\nu_{L\beta}\,,\nonumber \\
  \ell_{R\alpha} &\rightarrow& V_{\alpha\beta}\,\ell_{R\beta}\,,\ea
where $U$ and $V$ are arbitrary unitary matrices and $\alpha$ and
$\beta$ are the flavor indices. The mass term of charged leptons
and the effective mass term for Majorana neutrinos at the low
energy are of the form:\begin{equation}\label{Lagrangian}
    -\mathcal{L}=(m_{\ell})_{\alpha\beta}\overline{\ell_{R\alpha}}\ell_{L\beta}+\frac{1}{2}\,
    {(m_{\nu})}_{\alpha\beta}\overline{({\nu}_{L\alpha})^{c}}{\nu}_{L\beta}+\rm{H.c.} \end{equation}
In order for the Lagrangian to remain invariant under
transformations shown in Eq.~(\ref{transformation1}), the mass
matrices have to transform as follows
\begin{eqnarray}\label{transformation2}
m_{\ell} & \rightarrow & V\,m_{\ell}\,U^{\dagger}\,, \nonumber \\
m_{\nu} & \rightarrow &
U^{\ast}\,m_{\nu}\,U^{\dagger}\,.\end{eqnarray} (Notice that we
have here used the assumption that neutrinos are of Majorana
nature. For Dirac neutrinos with $\nu_{R \alpha} \to W_{\alpha
\beta} \nu_{R \beta}$, $m_\nu$ would transform as $W m_\nu
U^\dagger$ so all the following discussion should have been
reconsidered.) It can be readily shown that under transformations
in
Eq.~(\ref{transformation1})\begin{eqnarray}\label{transformation3}
m_{\nu}({m_{\ell}^{\dagger}}m_{\ell})^n & \rightarrow &
U^{\ast}\,m_{\nu}({m_{\ell}}^{\dagger}m_{\ell})^n\,U^{\dagger}\,,\nonumber\\
(m_{\ell}^{T}m_{\ell}^{\ast})^m m_{\nu} & \rightarrow &
U^{\ast}\,(m_{\ell}^{T}m_{\ell}^{\ast})^m m_{\nu}\,U^{\dagger}\,,
\end{eqnarray} where $m$ and $n$ are arbitrary integer numbers. In general,
a linear combination of these combinations also transforms in the
same way:\begin{equation}\label{combination}
\sum_{i}[a_{i}m_{\nu}({m_{\ell}^{\dagger}}m_{\ell})^{n_i}+b_{i}(m_{\ell}^{T}m_{\ell}^{\ast})^{m_i}
m_{\nu}]\;\rightarrow\;U^{\ast}\,\sum_{i}[a_{i}m_{\nu}({m_{\ell}^{\dagger}}m_{\ell})^{n_i}
+b_{i}(m_{\ell}^{T}m_{\ell}^{\ast})^{m_i}
m_{\nu}]\,U^{\dagger},\end{equation} where $a_{i}$ and $b_{i}$ are
arbitrary constants. In the above relation $m_i$ and $n_i$ can
take positive as well as negative integer numbers. Thus, the
determinant of this matrix will transform into itself times
$\mathrm{Det}[U^{\dagger}]\mathrm{Det}[U^{\ast}]$, which is a pure
phase. As a result, the ratio of any pair of such determinants is
invariant under the transformations shown in
Eq.~(\ref{transformation1}):\\
\begin{equation}\label{invariant}\frac{\mathrm{Det}[\sum_{i}
(a_{i}m_{\nu}({m_{\ell}^{\dagger}}m_{\ell})^{n_i}+b_{i}(m_{\ell}^{T}m_{\ell}^{\ast})^{m_i}
m_{\nu})]}{\mathrm{Det}[\sum_{i}(a_{i}'m_{\nu}({m_{\ell}^{\dagger}}m_{\ell})^{n^\prime_i}+b_{i}'(m_{\ell}^{T}m_{\ell}^{\ast})^{m^\prime_i}
m_{\nu})]}\;\ \mathrm{is}\;\ \mathrm{invariant.}\end{equation}
Moreover,
$$\mathrm{Det}\Big[\sum_{i}
(a_{i}m_{\nu}({m_{\ell}^{\dagger}}m_{\ell})^{n_i}+b_{i}(m_{\ell}^{T}m_{\ell}^{\ast})^{m_i}
m_{\nu})\Big]\left(\mathrm{Det}\Big[\sum_{i}(a_{i}'m_{\nu}({m_{\ell}^{\dagger}}m_{\ell})^{n^\prime_i}+b_{i}'(m_{\ell}^{T}m_{\ell}^{\ast})^{m^\prime_i}
m_{\nu})\Big]\right)^*$$ is also invariant.  Notice that
$m_\nu^n\equiv m_\nu (m_\nu^\dagger m_\nu)^{n-1}$ transforms
exactly in the same form as $m_\nu$  under basis transformations
({\it see}, Eqs.~(\ref{transformation1},\ref{transformation2})).
As a result, if we replace any of $m_\nu$ appearing in
Eq.~(\ref{invariant}) with $m_\nu^n$, the combination will
maintain its invariance under transformation
(\ref{transformation1}). The above combinations present an
infinite number of invariants. However the $3\times3$ Majorana
neutrino mass matrix contains nine degrees of freedom; so all of
these invariants cannot be independent. It is straightforward to
show that  there is a set of invariants which all the other
invariants can be written in terms of them. We will come back to
this point at the end of this section. In the following we give a
concrete example for such a ``complete'' set of invariants. We
will amply use these invariants in formulating the symmetries of
the neutrino mass matrix in sect.~\ref{projections}.

Let us define the following combination of mass matrices:
\begin{equation}\label{definitions}\mathcal{P}_1\equiv\frac{m_{\ell}^{-2}}
{tr[m_{\ell}^{-2}]}\, ,\;\;
\mathcal{P}_3\equiv\frac{{m_{\ell}^{\dagger}}m_{\ell}}{tr[{m_{\ell}^{\dagger}}m_{\ell}]}\,
, \;\;
\mathcal{P}_2\equiv\mathbf{I}-\mathcal{P}_1-\mathcal{P}_3\,,
\end{equation}where $\mathbf{I}$ is the three by three identity
matrix and $m_{\ell}^{-2}\equiv
m_{\ell}^{-1}({m_{\ell}^{-1}})^{\dagger}$. It is straightforward
to show that when the eigenvalues of $m_{\ell}^{\dagger}m_{\ell}$
are hierarchical, $\mathcal{P}_i$ act as projection operators. In
this section we will not use this property, but this feature will
play an important roll in the discussion of
sect.~\ref{projections}.

Now using these operators, let us define\\
\begin{equation}\mathcal{E}_1\equiv\frac{{\rm
Det}[(\mathcal{P}_2+\mathcal{P}_3)m_{\nu}-m_{\nu}\mathcal{P}_1]}{{\rm
Det}[m_{\nu}]},\nonumber \end{equation}
\begin{equation}\mathcal{E}_2\equiv\frac{{\rm
Det}[(\mathcal{P}_1+\mathcal{P}_3)m_{\nu}-m_{\nu}\mathcal{P}_2]}{{\rm
Det}[m_{\nu}]},\nonumber\end{equation}
\begin{equation}\label{definitionE}\mathcal{E}_3\equiv\frac{{\rm
Det}[(\mathcal{P}_1+\mathcal{P}_2)m_{\nu}-m_{\nu}\mathcal{P}_3]}{{\rm
Det}[m_{\nu}]},\end{equation} and
\begin{equation}\mathcal{F}_1\equiv\frac{{\rm
Det}[(\mathcal{P}_1+\mathcal{P}_3)m_{\nu}-m_{\nu}\mathcal{P}_3]}{{\rm
Det}[m_{\nu}]},\nonumber \end{equation}
\begin{equation}\mathcal{F}_2\equiv\frac{{\rm
Det}[(\mathcal{P}_2+\mathcal{P}_3)m_{\nu}-m_{\nu}\mathcal{P}_3]}{{\rm
Det}[m_{\nu}]},\nonumber\end{equation}
\begin{equation}\label{definitionF}\mathcal{F}_3\equiv\frac{{\rm
Det}[(\mathcal{P}_2+\mathcal{P}_3)m_{\nu}-m_{\nu}\mathcal{P}_2]}{{\rm
Det}[m_{\nu}]}.\end{equation}\\It is worth mentioning that only
{three} out of the six combinations defined in
Eqs.~(\ref{definitionE},\ref{definitionF}) are independent; that
is {three} of them can be written in terms of the other {three}
combinations. It can be shown that the combinations defined in
Eq.~(\ref{invariant}) can be in general written as a combination
of three $\mathcal{E}_i$ and $\mathcal{F}_i$ ({\it e.g.},
$\mathcal{E}_1$, $\mathcal{E}_2$ and $\mathcal{F}_1$).

Let us define invariants $\mathcal{G}_i$ and $\mathcal{H}_i$ by
replacing ${\rm Det}[m_{\nu}]$ in the denominators of
Eqs.~(\ref{definitionE},\ref{definitionF}) with ${\rm
Det}[m_{\nu}^3]$. That is
\begin{equation}\label{tarifGH}\mathcal{G}_i\equiv\frac{{\rm
Det}[m_{\nu}]}{{\rm Det}[m_{\nu}^3]}\;\mathcal{E}_i\quad\quad\quad
\text{and}\quad\quad\quad \mathcal{H}_i\equiv\frac{{\rm
Det}[m_{\nu}]}{{\rm Det}[m_{\nu}^3]}\;\mathcal{F}_i.\end{equation}
 It can be shown that four out of the six invariants $\mathcal{G}_i$
and $\mathcal{H}_i$ are independent.  Depending on the problem in
hand, one can choose four independent invariants  out of all these
twelve invariants to perform various analyses; {\it for example},
$\{ \mathcal{E}_1,\mathcal{E}_2,\mathcal{F}_1,\mathcal{G}_1 \}$.
An example of the application of such ``complete'' set of
invariants is formulating CP symmetry in the lepton sector. Both
in the quark and lepton sectors, using the Jarlskog invariant
($\mathcal{J}$) to test CP-violation is an established and widely
used technique \cite{Jarlskog}. As is well-known, the Majorana
neutrino mass matrix contains more than one source of CP-violation
and therefore more than one invariant will be necessary to check
CP-invariance \cite{kusenko,branco,us,sarkar}. Suppose we take a
``complete'' set of invariants for testing the CP-violation. If
any of these independent invariants is complex, the lepton sector
is not CP-invariant ({\it i.e.}, at least one of the three
possible CP-violating phases is different from zero). One may ask
whether the opposite is also correct. That is if all these
invariants are real, can we conclude that CP is conserved? Since
the equations are non-linear, if we take only three invariants, we
may in general find a specific solution in addition to the trivial
CP-invariant one. However, this specific solution may not be
compatible with the neutrino data. Once we examine the realness of
the fourth independent invariant, this solution can be excluded.

In the rest of this paper, we show how the combinations
$\mathcal{E}_i$ and $\mathcal{F}_i$ can facilitate the basis
independent analysis of the symmetries of neutrino sector.

\section{Symmetries of neutrino mass matrix in terms of the invariants\label{projections}}

There are conjectures (such as $\mu-\tau$ reflection and exchange
symmetries \cite{us,Harrison}, various flavor symmetries
\cite{Binetruy} and texture zero matrices for the neutrino mass
matrix \cite{glashow}) that are customized to be used for the
lepton sector. Any model that accommodates such conjectures also
has to reproduce the hierarchy of the charged lepton masses. In
this section, we use this property to formulate the conjectures in
a model-independent fashion.


In the charged lepton mass basis
 the operators $\mathcal{P}_1$,
 $\mathcal{P}_2$ and $\mathcal{P}_3$ take the following forms
\begin{equation}\label{definitions1}\mathcal{P}_1=\frac{1}{m_e^{-2}+m_{\mu}^{-2}+m_{\tau}^{-2}}
\left(
\begin{array}{ccc}
    m_e^{-2} & 0 & 0 \\
    0 & m_{\mu}^{-2} & 0 \\
    0 & 0 & m_{\tau}^{-2} \\
 \end{array} \right),\end{equation}
 \begin{equation}\label{definitions2}\mathcal{P}_3=\frac{1}{m_e^{2}+m_{\mu}^{2}+m_{\tau}^{2}}
\left(
\begin{array}{ccc}
    m_e^{2} & 0 & 0 \\
    0 & m_{\mu}^{2} & 0 \\
    0 & 0 & m_{\tau}^{2} \\
 \end{array} \right),\end{equation}and
 \begin{equation}\label{definitions3}\mathcal{P}_2=\left(
\begin{array}{ccc}
    1 & 0 & 0 \\
    0 & 1 & 0 \\
    0 & 0 & 1 \\
 \end{array} \right)-\mathcal{P}_1-\mathcal{P}_3.\end{equation}\\

Using  the hierarchy of the  charged lepton masses,
$$m_\mu^2/m_\tau^2=3.5 \times 10^{-3} \ll m_e^2/m_\mu^2= 2.3 \times
10^{-5}\ll m_e^2/m_\tau^2=7.9 \times 10^{-8},$$ we can write
\begin{equation}\label{tarifP1}\mathcal{P}_1= \left(
\begin{array}{ccc}
    1 & 0 & 0 \\
    0 & 0 & 0 \\
    0 & 0 & 0 \\
 \end{array}
 \right)+ {\cal {O}} \Big(\frac{m_{e}^2}{m_{\mu}^2},\alpha^2\Big),\end{equation}
 \begin{equation}\label{tarifP2}\mathcal{P}_2= \left(
\begin{array}{ccc}
    0 & 0 & 0 \\
    0 & 1-\alpha & 0 \\
    0 & 0 & \alpha \\
 \end{array}
 \right)+ {\cal {O}} \Big(\frac{m_{e}^2}{m_{\mu}^2},\alpha^2\Big),\end{equation}
 \begin{equation}\label{tarifP3}\mathcal{P}_3= \left(
\begin{array}{ccc}
    0 & 0 & 0 \\
    0 & \alpha & 0 \\
    0 & 0 & 1-\alpha \\
 \end{array}
 \right)+ {\cal {O}} \Big(\frac{m_{e}^2}{m_{\mu}^2},\alpha^2\Big),\end{equation} where $\alpha\equiv m_\mu^2/m_\tau^2$.

 In the charged lepton mass basis, to the first order of
the parameter $\alpha$,
  $\mathcal{E}_i$ and
$\mathcal{F}_i$ defined in
Eqs.~(\ref{definitionE},\ref{definitionF}) can be written as
\begin{equation}\mathcal{E}_1=\frac{m_{ee}(m_{\mu\tau}^{2}-m_{\mu\mu}m_{\tau\tau})}{{\rm
Det}[m_{\nu}]},\nonumber\end{equation}
\begin{equation}\mathcal{E}_2=\frac{m_{\mu\mu}(m_{e\tau}^{2}-m_{ee}m_{\tau\tau})}{{\rm
Det}[m_{\nu}]}(1-4\alpha),\nonumber\end{equation}
\begin{equation}\label{tarifE}\mathcal{E}_3=\frac{m_{\tau\tau}(m_{e\mu}^{2}-m_{ee}m_{\mu\mu})}{{\rm
Det}[m_{\nu}]}(1-4\alpha),\end{equation} and
\begin{equation}\mathcal{F}_1=\frac{m_{\mu\tau}(m_{ee}m_{\mu\tau}-m_{e\mu}m_{e\tau})}{{\rm
Det}[m_{\nu}]}(1-4\alpha)\nonumber,\end{equation}
\begin{equation}\mathcal{F}_2=\frac{m_{e\tau}(m_{\mu\mu}m_{e\tau}-m_{e\mu}m_{\mu\tau})}{{\rm
Det}[m_{\nu}]}(1-2\alpha),\nonumber\end{equation}
\begin{equation}\label{tarifF}\mathcal{F}_3=\frac{m_{e\mu}(m_{e\mu}m_{\tau\tau}-m_{e\tau}m_{\mu\tau})}{{\rm
Det}[m_{\nu}]}(1-2\alpha),\end{equation} where $m_{\alpha \beta}$
are the elements of $m_\nu$.
\\Notice that corrections
to these formulae comes from the next to leading order terms which
are of the order $\sim 10^{-5}$.

In the subsect.~\ref{mu-tau} we discuss  the $\mu-\tau$ symmetries
of the neutrino mass matrix and conserved $L_\mu-L_\tau$ flavor
symmetry in terms of the invariants. Then, in
subsect.~\ref{Naqsh-Sefr}, we discuss the so-called texture zero
matrices in an arbitrary basis for the lepton fields.

\subsection{$\mu-\tau$ symmetries of $m_\nu$\label{mu-tau}}

In this subsection, we study the $\mu-\tau$ reflection and
exchange symmetries by using $\mathcal{E}_i$ and $\mathcal{F}_i $.
The aim is to formulate these symmetries in a basis independent
way. We first study the $\mu-\tau$ exchange symmetry, and  we then
turn our attention to the $\mu-\tau$ reflection
symmetry \cite{us,Harrison}. \\
The $\mu-\tau$ exchange symmetry (the symmetry under
$\nu_\mu\leftrightarrow\nu_\tau\,$)
implies\begin{equation}\label{exchange}m_{e\mu}=m_{e\tau}\quad,\quad
m_{\mu\mu}=m_{\tau\tau},\end{equation} where $m_{\alpha\beta}$ are
the entries of $m_\nu$ in the charged lepton mass basis. From Eqs.
(\ref{tarifE},\ref{tarifF}), these conditions can be expressed in
terms of $\mathcal{E}_i$ and $\mathcal{F}_i$ in the following
form\\
\begin{equation}\label{exchange1}\left|\frac{\mathcal{F}_2-\mathcal{F}_3}
{\mathcal{F}_2+\mathcal{F}_3}\right|\ll1 \ \ {\rm or \
equivalently} \ \ \left|\frac{\mathcal{E}_2-\mathcal{E}_3}
{\mathcal{E}_2+\mathcal{E}_3}\right|\ll1.\end{equation}\\
The above inequalities are basis independent criteria for the
$\mu-\tau$ exchange symmetry. In any basis to check for the
$\mu-\tau$ exchange symmetry, we can immediately compute the
combination in the Eq.~(\ref{exchange1}); if  the condition in
this equation is satisfied, the neutrino mass matrix is symmetric
under the $\mu-\tau$ exchange.

Now let us consider the $\mu-\tau$ reflection symmetry. In the
mass basis of charged leptons, the symmetry under
$\nu_\mu\leftrightarrow\nu_\tau^\ast\,$, (with proper rephasing)
implies\begin{equation}\label{reflection}
m_{e\mu}=m_{e\tau}^\ast\quad,\quad
m_{\mu\mu}=m_{\tau\tau}^\ast\quad,\quad
m_{ee}=m_{ee}^\ast\quad,\quad
m_{\mu\tau}=m_{\mu\tau}^\ast.\end{equation} It is straightforward
to show that these equalities, expressed in terms of
$\mathcal{E}_i$ and $\mathcal{F}_i$, implies the following
inequalities\begin{equation}\label{reflection1}\left|\frac{\mathcal{F}_3-\mathcal{F}_2^*}
{\mathcal{F}_3+\mathcal{F}_2^*}\right|\ll1,\;\;\quad\quad\text{and}\;\;\quad\quad
\left|\frac{\mathcal{E}_1-\mathcal{E}_1^*}
{\mathcal{E}_1+\mathcal{E}_1^*}\right|\ll1.\end{equation}

As discussed in sect. \ref{tarif}, the above ratios are invariant
under general basis transformation Eq.~(\ref{transformation1}).
Thus, we have found some basis independent criteria for testing
the $\mu-\tau$ reflection symmetry; if either of the inequalities
in Eq.~(\ref{reflection1}) does not hold in a given basis, the
neutrino mass matrix is not symmetric under the $\mu-\tau$
reflection.

The relations in Eqs.~(\ref{exchange1}) and (\ref{reflection1})
are the necessary but {\it not} sufficient conditions for
$\mu-\tau$ exchange symmetry and $\mu-\tau$ reflection symmetry,
respectively. If the inequalities in Eqs.~(\ref{exchange1}) or
(\ref{reflection1}) do not hold in a given basis, we can conclude
that lepton sector is not symmetric under the corresponding
transformations. But the reverse is not correct; that is in
certain very special cases it is possible to satisfy the
conditions in Eqs.~(\ref{exchange1}) or (\ref{reflection1})
without having the corresponding symmetry.

Another symmetry proposed in the literature, which has some common
features with the $\mu-\tau$ exchange symmetry, is the flavor
symmetry with conserved $L_\mu-L_\tau$ \cite{Binetruy}. In the
charged lepton mass basis, the conservation of $L_\mu-L_\tau$
implies the following form for the neutrino mass matrix
\begin{equation}\label{lmu-ltau}\left(
\begin{array}{ccc}
    \times & 0 & 0 \\
    0 & 0 & \times \\
    0 & \times & 0 \\
 \end{array}
 \right),\end{equation}
where ``$\times$'' means that the corresponding entry is nonzero.
In the limit of approximation made in Eqs.~(\ref{tarifE}) and
(\ref{tarifF}), for a mass matrix of the form
Eq.~(\ref{lmu-ltau}), the invariants $\mathcal{E}_i$ and
$\mathcal{F}_i$ take the following values
\begin{equation}\label{lmu-ltau1} \mathcal{E}_2=
\mathcal{E}_3=\mathcal{F}_2=\mathcal{F}_3=0,\quad\;
\mathcal{E}_1=-1\quad\; \text{and}\quad\;
\mathcal{F}_1=-1+\alpha.\end{equation} [Corrections to these
values are of the order $\sim {\cal {O}}(\alpha^2\sim 10^{-5})$.]
The above criterion for the flavor symmetry conserving
$L_\mu-L_\tau$, is a necessary but {\it not} a sufficient
criterion for this symmetry. That is, if a model does not satisfy
this criterion in one basis, the model does not respect the
$L_\mu-L_\tau$ conserving flavor symmetry; but the reverse is not
correct.

Notice that a mass matrix of form Eq.~(\ref{lmu-ltau}) cannot
accommodate the present data of neutrino experiments because it
predicts two degenerate mass eigenvalues. In order to accommodate
the present data, the $L_\mu-L_\tau$ conserving symmetry has to be
broken; {\it i.e.}, $L_\mu-L_\tau$ can be only an approximate
symmetry. Thus, in practice the exact equalities, ``$=$'', in
Eq.~(\ref{lmu-ltau1}) has to be replaced by ``$\simeq$''.

\subsection{Texture zero matrices in terms of
invariants\label{Naqsh-Sefr}}

Among the various conjectures that can be imposed on the neutrino
mass matrix, the texture zero ansatz have received more attention
in the literature \cite{glashow,Xing}. In these scenarios, certain
entries of the neutrino mass matrix in the charged lepton mass
basis are conjectured to be equal to zero. Such an assumption can
originate from a more fundamental theory or an underlying symmetry
\cite{texture}. On the other hand, most of the models of the
neutrino mass matrix are built based on some symmetry that is
apparent only in a certain basis which may not correspond to the
mass basis of the charged leptons. In this section, we show that
by using the combinations $\mathcal{E}_i$ and $\mathcal{F}_i$
defined in Eqs. (\ref{definitionE},\ref{definitionF}), we can
express the conditions for texture zero in a basis-independent
way.

 As shown in \cite{glashow}, a neutrino mass matrix $m_\nu$ with three or more zero
entries is not compatible with the data. Moreover, among the
fifteen possible two zero texture mass matrices, only seven of
them can be made compatible with the present neutrino data
\cite{glashow}. The two zero textures are labeled $A_1$, $A_2$,
$B_1$, $B_2$, $B_3$, $B_4$ and $C$.
 We have listed them in
Table \ref{20texture}. The non-vanishing entries in this Table are
denoted by $\times$. These textures have certain predictions for
the values of the neutrino parameters. For example, the $A_1$ and
$A_2$ textures are compatible with data only for normal
hierarchical scheme ($m_1\ll \sqrt{\Delta m_{atm}^2}$).

From the Eqs.~(\ref{tarifE},\ref{tarifF}), we readily observe that
if one of the diagonal or off-diagonal elements of the mass matrix
vanishes, some of $\mathcal{E}_i$ or $\mathcal{F}_i$ will also go
to zero ({\it e.g.,} $m_{ee}=0 \Longrightarrow \mathcal{E}_1\to
0;m_{\mu \tau}=0 \Longrightarrow \mathcal{F}_1\to 0$ and so on).
Remember that under basis transformation, $\mathcal{F}_i$ and
$\mathcal{E}_i$ are invariant. Thus, if an $\mathcal{E}_i$ or an
$\mathcal{F}_i$ vanishes in a particular basis, it will vanish in
all bases. To be precise, there is a correction of order of
$\alpha^2\sim m_e^2/m_\mu^2\sim 10^{-5}$ to $\mathcal{E}_i$ and
$\mathcal{F}_i$ shown in Eqs. (\ref{tarifE},\ref{tarifF}). As a
result, when a mass matrix element vanishes,  certain
$\mathcal{F}_i$ and $\mathcal{E}_i$ become much smaller than the
rest but not exactly zero.

\begin{table}[!t]
  \centering
  \caption{Two zero texture mass matrices}\label{20texture}
  \vspace{0.4in}
  \begin{tabular}{l r}
    Label & Neutrino mass matrix $m_\nu$\\[2mm]
    \hline\\
    $A_1$ \hspace{1.5in} & $ \left(
\begin{array}{ccc}
    0 & 0 & \times \\
    0 & \times & \times \\
    \times & \times & \times \\
 \end{array} \right) $ \\[7mm]
    \hline\\
    $A_2$ & $\left(
\begin{array}{ccc}
    0 & \times & 0 \\
    \times & \times & \times \\
    0 & \times & \times \\
 \end{array} \right) $ \\[7mm]
 \hline\\
    $B_1$ & $\left(
\begin{array}{ccc}
    \times & \times & 0 \\
    \times & 0 & \times \\
    0 & \times & \times \\
 \end{array} \right) $ \\[7mm]
 \hline\\
    $B_2$ & $\left(
\begin{array}{ccc}
    \times & 0 & \times \\
    0 & \times & \times \\
    \times & \times & 0 \\
 \end{array} \right) $ \\[7mm]
 \hline\\
    $B_3$ & $\left(
\begin{array}{ccc}
    \times & 0 & \times \\
    0 & 0 & \times \\
    \times & \times & \times \\
 \end{array} \right) $ \\[7mm]
 \hline\\\
    $B_4$ & $\left(
\begin{array}{ccc}
    \times & \times & 0 \\
    \times & \times & \times \\
    0 & \times & 0 \\
 \end{array} \right) $ \\[7mm]
 \hline\\
    $C$ & $\left(
\begin{array}{ccc}
    \times & \times & \times \\
    \times & 0 & \times \\
    \times & \times & 0 \\
 \end{array} \right) $ \\[7mm]
    \hline
  \end{tabular}
\end{table}
\newpage
\begin{table}[!t]
  \centering
  \caption{Values of $\mathcal{E}_i$ and $\mathcal{F}_i$ (leading order) for textures in Table \ref{20texture}}\label{values}
  \vspace{0.4in}
  \begin{tabular}{l l}
  Label & Values of $\mathcal{E}_i$ and $\mathcal{F}_i$\\[3mm]
  \hline\\
  $A_1$ \hspace{1.5in} & $\mathcal{E}_1\approx \mathcal{E}_3\approx\mathcal{F}_1\approx
  \mathcal{F}_3\ll \mathcal{E}_2\approx\mathcal{F}_2$
  \\[3mm]
  \hline
  \\
  $A_2$ & $\mathcal{E}_1\approx\mathcal{E}_2\approx\mathcal{F}_1\approx\mathcal{F}_2
  \ll\mathcal{E}_3\approx\mathcal{F}_3$ \\[3mm]
  \hline \\
  $B_1$ & $\mathcal{E}_2\approx\mathcal{F}_2\ll \mathcal{E}_1\approx\mathcal{F}_1,
  \mathcal{E}_3\approx\mathcal{F}_3$\\[3mm]
  \hline\\
  $B_2$ & $\mathcal{E}_3\approx\mathcal{F}_3\ll\mathcal{E}_1\approx\mathcal{F}_1,
  \mathcal{E}_2\approx\mathcal{F}_2$\\[3mm]
  \hline\\
  $B_3$ & $\mathcal{E}_2\approx\mathcal{E}_3\approx\mathcal{F}_2\approx\mathcal{F}_3\ll
  \mathcal{E}_1\approx\mathcal{F}_1$ \\[3mm]
  \hline\\
  $B_4$ & $\mathcal{E}_2\approx\mathcal{E}_3\approx\mathcal{F}_2\approx\mathcal{F}_3\ll
  \mathcal{E}_1\approx\mathcal{F}_1$ \\[3mm]
   \hline\\
  $C$ & $\mathcal{E}_2\approx\mathcal{E}_3\ll \mathcal{F}_2\approx\mathcal{F}_3\approx
  \mathcal{F}_1-\mathcal{E}_1\neq0$  \\[3mm]
  \hline
  \end{tabular}
  \vspace{1.0in}
\end{table}
 Each of
the texture zeros implies a certain pattern for $\mathcal{E}_i$
and $\mathcal{F}_i$. For example, for the $A_1$ texture we find
$\mathcal{E}_1\approx\mathcal{E}_3\approx\mathcal{F}_1\approx\mathcal{F}_3\ll
\mathcal{E}_2\approx\mathcal{F}_2$. The patterns of
$\mathcal{E}_i$ and $\mathcal{F}_i$ for the rest of the textures
are summarized in Table \ref{values}. Here, $A\approx B$ means
$|A-B|/|A+B|\lesssim O(\alpha^2\sim 10^{-5})$. Going to higher
orders of $\alpha$ makes the analysis {\it uselessly} cumbersome,
especially that  in most models that predict texture zeros the
vanishing elements of $m_\nu$ receive a small correction (due to
running or etc.).
 Thus, in the following we consider
the leading order patterns for $\mathcal{E}_i$ and
$\mathcal{F}_i$. That is to perform the analysis, we will replace
$``\approx"$ with $``="$ and set $\mathcal{E}_i$ and
$\mathcal{F}_i$ for each pattern that according to Table
\ref{values} are much smaller than the rest equal to zero.

These patterns can be considered as a test for the two zero
textures. That is, by computing $\mathcal{E}_i$ and
$\mathcal{F}_i$ in any given basis, one can check if a certain
pattern can be the case. It is obvious that if the pattern
associated with a certain texture does not hold, $m_\nu$ in the
charged lepton mass basis will not have the format of that
particular texture. In the following, we explore whether the
opposite is also true. The question is as follows. Suppose that a
certain pattern of $\mathcal{E}_i$ and $\mathcal{F}_i$ listed in
Table \ref{values} is realized. Can we then conclude that  $m_\nu$
in the charged lepton mass basis has the format of the texture
corresponding to that particular pattern? To answer this question
we check if the equations listed in Table \ref{values} have any
solution compatible with the neutrino data other than the
particular texture zero solution corresponding to them. To perform
the analysis we use the standard parametrization of the neutrino
mass matrix presented by the particle data group \cite{pdg}.

Let us first discuss  the $A_1$ and $A_2$ textures. Notice that
among the textures listed in Table \ref{values}, only for the
$A_1$ and $A_2$  textures we have $\mathcal{F}_1=\mathcal{E}_1=0$.
In the following, we first check if, despite $m_{ee},m_{\mu \tau}
\ne 0$, we can have $\mathcal{F}_1=\mathcal{E}_1=0$ and then check
for solutions with $m_{ee}=0,m_{\mu \tau}\ne 0$ and $m_{ee}\ne 0,
m_{\mu \tau}=0$. It is straightforward but rather cumbersome to
show that, assuming $m_{ee},m_{\mu \tau} \ne 0$, the  only
solution of $\mathcal{F}_1=\mathcal{E}_1=0$ is $m_3=0, s_{13}=0$.
(In fact, there is another solution which requires
$m_1=-m_2(c_{12}^2-c_{12}s_{12}s_{13} \tan \theta_{23}
e^{i\delta})/(s_{12}^2+s_{12}c_{12}s_{13}\tan \theta_{23} e^{i
\delta})$ but this is not compatible with neutrino data.) It is
straightforward to show that $m_3=s_{13}=0$ implies
$\mathcal{E}_2,\mathcal{E}_3 \ne 0$ so not all of the conditions
for the $A_1$ and $A_2$ textures can be fulfilled. Thus, so far we
have concluded that if the pattern associated to the $A_1$ or
$A_2$ textures holds (if
$\mathcal{E}_1=\mathcal{E}_3=\mathcal{F}_1=\mathcal{F}_3=0$ or
$\mathcal{E}_1=\mathcal{E}_2=\mathcal{F}_1=\mathcal{F}_2=0$) at
least one of the ${ee}$ or ${\mu \tau}$ entries must be nonzero.
On the other hand, $m_{ee}=0$ and $\mathcal{E}_1=\mathcal{F}_1=0$
with $m_{\mu \tau} \ne 0$ implies $m_{e\mu}m_{e\tau}=0$ which is
the condition for the $A_1$ or $A_2$ textures. These two textures
can be distinguished by computing $\mathcal{E}_2$ and
$\mathcal{E}_3$ and checking which one vanishes. Finally,
$m_{ee}\ne 0$, $m_{\mu \tau}=0$ and
$\mathcal{E}_1=\mathcal{F}_1=0$ implies $m_{\mu \mu} m_{\tau
\tau}=0$ which is not compatible with the data \cite{glashow}. In
sum, we have proved that the equations
$\mathcal{E}_1=\mathcal{E}_3=\mathcal{F}_1=\mathcal{F}_3=0$
($\mathcal{E}_1=\mathcal{E}_2=\mathcal{F}_1=\mathcal{F}_2=0$) are
both necessary and sufficient conditions for the $A_1$ ($A_2$)
texture provided that the mass matrix accommodates the present
neutrino data.

Now let us discuss the $B_i$ textures. As shown in Table
\ref{values}, the conditions for $B_1$ are
$\mathcal{E}_2=\mathcal{F}_2=0$, $\mathcal{E}_1=\mathcal{F}_1$ and
$\mathcal{E}_3=\mathcal{F}_3$.  Notice that
$\mathcal{E}_2=\mathcal{F}_2=0$ automatically implies
$\mathcal{E}_1=\mathcal{F}_1$ and $\mathcal{E}_3=\mathcal{F}_3$.
Thus to check if the conditions shown in the third row of Table
\ref{values}  guarantee the format of texture $B_1$, it is
sufficient to solve $\mathcal{E}_2=\mathcal{F}_2=0$. It can be
shown that $\mathcal{E}_2=\mathcal{F}_2=0$ (the conditions for
$B_1$), in addition to $m_{\mu \mu}=m_{e\tau}=0$, have another
solution which yields the following relations \ba \left\{
\begin{matrix}|m_2|^2-|m_1|^2 &=& |m_1|^2\cos
\delta (4 s_{13})/({s_{12}c_{12}})\cr |m_3|^2-|m_1|^2 &=&|m_1|^2
(s_{23}^2-c_{23}^2)/c_{23}^4\end{matrix}\right. \label{relation1}\
. \ea As a result, $s_{13} \cos \delta/(s_{23}^2-c_{23}^2)\sim
(\Delta m_{sol}^2)/(\Delta m_{atm}^2) \ll 1 \ . $  Moreover, the
second equation of Eq.~(\ref{relation1}) can be considered as a
lower bound on $|m_1|$; with the present data \cite{Schwetz}, this
equation gives the bound  $|m_1|>0.006$~eV at 3$\sigma$. Notice
that with the present uncertainties on the neutrino data this
solution is still acceptable. Thus, the conditions listed in the
third row of Table \ref{values}, in addition to texture $B_1$ have
another solution which is compatible with the present neutrino
data. Future measurements of the neutrino mass scale
\cite{majorana-phases}, $\theta_{23}$ and ${\rm
sgn}(|m_3|^2-|m_1|^2)$ may enable us to test the second equation
in Eq. (\ref{relation1}). In particular, the NO$\nu$A \cite{nova}
and T2K \cite{T2K} experiments can measure $\theta_{23}$ and the
absolute value of $\Delta m_{31}^2$ with very high accuracy. The
accuracy in the measurement of $\sin^2 2 \theta_{23}$ can reach
1\%. If these experiments establish that $\theta_{23}$ is close to
maximal, the lower bound on $|m_1|$ will be within the reach of
the KATRIN experiment \cite{KATRIN}. For relatively large values
of $\theta_{13}$ ({\it i.e.,} $s_{13}>0.05$), more futuristic
experiments such as the T2KK setup \cite{T2KK} can help us to
solve the octant-degeneracy and derive information on ${\rm
sgn}(|m_3|^2-|m_1|^2)$ and $\delta$. Such information makes the
solution completely testable. In summary, the conditions listed in
the third row of Table \ref{values}, in addition to texture $B_1$
have another solution compatible with the present data.
Forthcoming data may exclude this solution.


Now let us study the conditions for texture $B_2$ which are listed
in the fourth row of Table \ref{values}. The conditions
$\mathcal{E}_3=\mathcal{F}_3=0$  automatically yield
$\mathcal{E}_1=\mathcal{F}_1$ and $\mathcal{E}_2 =\mathcal{F}_2$
so it will be sufficient to study the consequences of
$\mathcal{E}_3=\mathcal{F}_3=0$. Similarly to the case of texture
$B_1$, $\mathcal{E}_3=\mathcal{F}_3=0$ for $m_{e \mu}, m_{\tau
\tau}\ne 0$ implies \ba \left\{
\begin{matrix}|m_2|^2-|m_1|^2 &=& -|m_1|^2\cos
\delta (4 s_{13})/({s_{12}c_{12}})\cr |m_3|^2-|m_1|^2 &=&-|m_1|^2
(s_{23}^2-c_{23}^2)/c_{23}^4\end{matrix}\right. \label{relation2}\
. \ea A discussion similar to the one after Eqs.~(\ref{relation1})
holds here, too. That is, the conditions
$\mathcal{E}_3=\mathcal{F}_3=0$ other than texture $B_2$ has
another solution which is compatible with the present data but can
be excluded by the forthcoming NO$\nu$A \cite{nova} and T2K
\cite{T2K} experiments.

Now let us discuss the $B_3$ and $B_4$ textures whose necessary
conditions are
$\mathcal{E}_2=\mathcal{E}_3=\mathcal{F}_2=\mathcal{F}_3=0$. From
Eqs. (\ref{relation1},\ref{relation2}), we readily see that if
$\mathcal{E}_2=\mathcal{E}_3=\mathcal{F}_2=\mathcal{F}_3=0$, there
is no solution with $m_{e\mu},m_{e\tau},m_{\mu \mu},m_{\tau \tau}
\ne 0$. That is some of these entries should vanish. Considering
the different configurations of vanishing  entries, we find that
$\mathcal{E}_2=\mathcal{F}_2=\mathcal{E}_3=\mathcal{F}_3=0$
implies either  $B_3$ or $B_4$.

Finally, let us discuss the condition for texture $C$ whose
conditions are listed in the last row of Table~\ref{values}.
Notice that for $m_{\mu \mu},m_{\tau \tau}\ne 0$,
$\mathcal{E}_2=\mathcal{E}_3=0$ automatically yields
$\mathcal{F}_2=\mathcal{F}_3=\mathcal{F}_1-\mathcal{E}_1$. In
addition to $m_{\mu \mu}=m_{\tau \tau}=0$, the equations
$\mathcal{E}_2=\mathcal{E}_3=0$ have another solution which
implies \ba
\begin{matrix}{|m_2|^2-|m_1|^2 \over |m_1|^2}&=& {\cos 2
\theta_{12} \over s_{12}^4}\,,\cr {|m_3|^2-|m_1|^2 \over |m_1|^2}
&\simeq& {\cos^2 (2\theta_{23}) c_{12}^2 \over 4s_{23}^4c_{23}^2
s_{13}^2 s_{12}^2}\,.\end{matrix} \  \label{relation3} \ea
 The above relations  in turn implies $$ \Delta m_{atm}^2 /\Delta
m_{sol}^2= (s_{12}^2 c_{12}^2/\cos 2\theta_{12})(1
/4s_{23}^4c_{23}^2)[\cos^2 (2\theta_{23}) /s_{13}^2]\gg 1.$$ The
above relation might be tested by forthcoming measurements.
If these experiments do not confirm Eq.~(\ref{relation3}), the
aforementioned solution will be ruled out and the texture $C$ will
be the only solution of $\mathcal{E}_2=\mathcal{E}_3=0$.

In sum, we have listed the necessary conditions for different
texture zero scenarios  in Table \ref{values}. We have shown that
in the case of texture $A_1$ and $A_2$ the conditions listed
respectively in the first and second rows of this Table are
sufficient to establish  these textures. Moreover, in the case of
textures $B_3$ and $B_4$, the conditions listed in the Table have
no solution compatible with the neutrino data other than these
textures. However, the conditions for $B_1, \ B_2 $ and $C$ can
have another solution which might be ruled out by improving the
neutrino data.

\section{Summary \label{summary}}
In this paper, we have studied the symmetries of the lepton sector
in a basis independent way by defining a new class of basis
invariants constructed out of the lepton mass matrices. We have
focused on the symmetries of the effective mass matrix of
neutrinos at low energies (below the electroweak scale) under the
assumption that neutrinos are Majorana particles.


As is well-known, even in the most optimistic case through the
present methods and proposals, the neutrino mass matrix cannot be
fully reconstructed. Motivated by this observation, various
neutrino mass matrix ansatz have been developed in the literature.
Most of these conjectures are based on symmetries and conditions
that are apparent only in a particular basis. We have shown that
by using the invariants defined in this paper such symmetries and
conditions can be formulated in a basis independent way. We have
in particular focused on the $\mu-\tau$ exchange and reflection
symmetries, flavor symmetry with conserved $L_\mu-L_\tau$; and
texture zeros. We have demonstrated how our invariants can
facilitate testing the conditions defining these ansatz in a
general basis.

\section*{Acknowledgment}
The authors are  grateful to M. M. Sheikh-Jabbari for careful
reading of the manuscript and the useful discussions.


\begin{thebibliography}{10}

\bibitem{global}
 S.~M.~Bilenky,
  Prog.\ Part.\ Nucl.\ Phys.\  {\bf 57}, 61 (2006)
  [arXiv:hep-ph/0510175];
 G.~L.~Fogli, E.~Lisi, A.~Marrone and A.~Palazzo,
  Prog.\ Part.\ Nucl.\ Phys.\  {\bf 57}, 742 (2006)
  [arXiv:hep-ph/0506083].

\bibitem{kamland}
  F.~Suekane  [KamLAND Collaboration],
  Prog.\ Part.\ Nucl.\ Phys.\  {\bf 57}, 106 (2006).

\bibitem{kek}
  E.~Aliu {\it et al.}  [K2K Collaboration],
  Phys.\ Rev.\ Lett.\  {\bf 94}, 081802 (2005)
  [arXiv:hep-ex/0411038];
 M.~Bishai  [MINOS Collaboration],
  AIP Conf.\ Proc.\  {\bf 903}, 271 (2007).

\bibitem{majorana-phases}
 Y.~Farzan, O.~L.~G.~Peres and A.~Yu.~Smirnov,
  Nucl.\ Phys.\  B {\bf 612}, 59 (2001)
  [arXiv:hep-ph/0105105];
S.~M.~Bilenky, S.~Pascoli and S.~T.~Petcov,
  Phys.\ Rev.\  D {\bf 64}, 053010 (2001)
  [arXiv:hep-ph/0102265];
V.~Barger, S.~L.~Glashow, P.~Langacker and D.~Marfatia,
  Phys.\ Lett.\  B {\bf 540}, 247 (2002)
  [arXiv:hep-ph/0205290];
  S.~Pascoli, S.~T.~Petcov and T.~Schwetz,
  Nucl.\ Phys.\  B {\bf 734}, 24 (2006)
  [arXiv:hep-ph/0505226];
  A.~Joniec and M.~Zralek,
  Phys.\ Rev.\  D {\bf 73}, 033001 (2006)
  [arXiv:hep-ph/0411070];
  S.~Pascoli, S.~T.~Petcov and W.~Rodejohann,
  Phys.\ Lett.\  B {\bf 549}, 177 (2002)
  [arXiv:hep-ph/0209059];
  H.~Nunokawa, W.~J.~C.~Teves and R.~Zukanovich Funchal,
  Phys.\ Rev.\  D {\bf 66}, 093010 (2002)
  [arXiv:hep-ph/0206137].

\bibitem{Jarlskog}
  C.~Jarlskog,
  Phys.\ Rev.\ Lett.\  {\bf 55}, 1039 (1985);
  C.~Jarlskog,
  Z.\ Phys.\  C {\bf 29}, 491 (1985).


\bibitem{kusenko}
  A.~Kusenko and R.~Shrock,
  [arXiv:hep-ph/9403315];
  P.~J.~O'Donnell and U.~Sarkar,
  Phys.\ Rev.\  D {\bf 52}, 1720 (1995)
  [arXiv:hep-ph/9305338];
 H.~K.~Dreiner, J.~S.~Kim, O.~Lebedev and M.~Thormeier,
  Phys.\ Rev.\  D {\bf 76}, 015006 (2007)
  [arXiv:hep-ph/0703074];
  J.~F.~Nieves and P.~B.~Pal,
  Phys.\ Rev.\  D {\bf 64}, 076005 (2001)
  [arXiv:hep-ph/0105305].

\bibitem{branco}
  G.~C.~Branco and M.~N.~Rebelo,
  New J.\ Phys.\  {\bf 7}, 86 (2005)
  [arXiv:hep-ph/0411196].


\bibitem{us}
 Y.~Farzan and A.~Yu.~Smirnov,
  JHEP {\bf 0701}, 059 (2007)
  [arXiv:hep-ph/0610337].

\bibitem{sarkar}
  U.~Sarkar and S.~K.~Singh,
  Nucl.\ Phys.\  B {\bf 771}, 28 (2007)
  [arXiv:hep-ph/0608030].


\bibitem{branco1}
  G.~C.~Branco, L.~Lavoura and M.~N.~Rebelo,
  Phys.\ Lett.\  B {\bf 180} (1986) 264.


\bibitem{Harrison}
  P.~F.~Harrison and W.~G.~Scott,
  Phys.\ Lett.\  B {\bf 547}, 219 (2002)
  [arXiv:hep-ph/0210197];
  W.~Grimus and L.~Lavoura,
  Phys.\ Lett.\  B {\bf 579}, 113 (2004)
  [arXiv:hep-ph/0305309].



\bibitem{Binetruy}
  P.~Binetruy, S.~Lavignac, S.~T.~Petcov and P.~Ramond,
  Nucl.\ Phys.\  B {\bf 496} (1997) 3
  [arXiv:hep-ph/9610481];
  N.~F.~Bell and R.~R.~Volkas,
  Phys.\ Rev.\  D {\bf 63} (2001) 013006
  [arXiv:hep-ph/0008177];
  K.~S.~Babu, E.~Ma and J.~W.~F.~Valle,
  Phys.\ Lett.\  B {\bf 552} (2003) 207
  [arXiv:hep-ph/0206292];
  S.~Choubey and W.~Rodejohann,
  Eur.\ Phys.\ J.\  C {\bf 40} (2005) 259
  [arXiv:hep-ph/0411190];
  W.~Rodejohann and M.~A.~Schmidt,
  Phys.\ Atom.\ Nucl.\  {\bf 69} (2006) 1833
  [arXiv:hep-ph/0507300].

\bibitem{glashow}
  P.~H.~Frampton, S.~L.~Glashow and D.~Marfatia,
  Phys.\ Lett.\  B {\bf 536}, 79 (2002)
  [arXiv:hep-ph/0201008].

\bibitem{Xing}
  Z.~z.~Xing,
  Int.\ J.\ Mod.\ Phys.\  A {\bf 19}, 1 (2004)
  [arXiv:hep-ph/0307359];
  S.~Zhou and Z.~z.~Xing,
  Eur.\ Phys.\ J.\  C {\bf 38}, 495 (2005)
  [arXiv:hep-ph/0404188];
  S.~Dev, S.~Kumar, S.~Verma and S.~Gupta,
  [arXiv:hep-ph/0708.3321];
  S.~Kaneko, H.~Sawanaka and M.~Tanimoto,
  JHEP {\bf 0508}, 073 (2005)
  [arXiv:hep-ph/0504074].

\bibitem{texture}
  S.~Kaneko, H.~Sawanaka, T.~Shingai, M.~Tanimoto and K.~Yoshioka,
  [arXiv:hep-ph/0703250];
  W.~Grimus,
  PoS {\bf HEP2005}, 186 (2006)
  [arXiv:hep-ph/0511078];
  W.~Grimus, A.~S.~Joshipura, L.~Lavoura and M.~Tanimoto,
  Eur.\ Phys.\ J.\  C {\bf 36}, 227 (2004)
  [arXiv:hep-ph/0405016];
  L.~Lavoura,
  Phys.\ Lett.\  B {\bf 609}, 317 (2005)
  [arXiv:hep-ph/0411232];
  P.~H.~Frampton, M.~C.~Oh and T.~Yoshikawa,
  Phys.\ Rev.\  D {\bf 66}, 033007 (2002)
  [arXiv:hep-ph/0204273];
  M.~S.~Berger and K.~Siyeon,
  Phys.\ Rev.\  D {\bf 64}, 053006 (2001)
  [arXiv:hep-ph/0005249].



\bibitem{pdg}
  W.~M.~Yao {\it et al.}  [Particle Data Group],
  J.\ Phys.\ G {\bf 33}, 1 (2006).




\bibitem{Schwetz}
  T.~Schwetz, M.~Tortola and J.~W.~F.~Valle,
  arXiv:0808.2016 [hep-ph];
  A.~Strumia and F.~Vissani,
  Nucl.\ Phys.\  B {\bf 726}, 294 (2005)
  [arXiv:hep-ph/0503246].



\bibitem{nova}
  D.~S.~Ayres {\it et al.}  [NOvA Collaboration],
  [arXiv:hep-ex/0503053].

\bibitem{T2K}
  Y.~Itow {\it et al.}  [The T2K Collaboration],
  [arXiv:hep-ex/0106019].

\bibitem{KATRIN}
  A.~Osipowicz {\it et al.}  [KATRIN Collaboration],
  [arXiv:hep-ex/0109033].

\bibitem{T2KK}
  K.~Hagiwara and N.~Okamura,
  [arXiv:hep-ph/0611058];
  K.~Hagiwara, N.~Okamura and K.~i.~Senda,
  [arXiv:hep-ph/0607255];
  T.~Kajita, H.~Minakata, S.~Nakayama and H.~Nunokawa,
  Phys.\ Rev.\  D {\bf 75}, 013006 (2007)
  [arXiv:hep-ph/0609286].




\end{thebibliography}
\end{document}